\documentclass[12pt]{article}
\usepackage{amsmath}
\usepackage{amsfonts}
\usepackage{amssymb}
\usepackage{xcolor}
\usepackage{color}
\usepackage[english]{babel}
\usepackage{graphics}
\usepackage{graphicx}
\begin{document}
\title{Multiverse as an ensemble of stable and unstable Universes}
\author{K. Urbanowski\footnote{e--mail:  K.Urbanowski@if.uz.zgora.pl}, \\
\hfill\\
University of Zielona G\'{o}ra, Institute of Physics, \\
ul. Prof. Z. Szafrana 4a, 65--516 Zielona G\'{o}ra, Poland.
}
\maketitle

\begin{abstract}

 Estimates of the Higgs and top quark masses, $m_{H} \simeq 125.10 \pm 0.14$ [GeV] and $m_{t} \simeq 172.76 \pm 0.30$ [GeV]  based on the experimental results  place the Standard Model in the region of the  metastable vacuum.
A consequence of the  metastability of the Higgs vacuum  is that it should
induce the decay of the electroweak vacuum in the early Universe with catastrophic consequences.
It may happen that certain universes were lucky enough to survive the time of canonical decay, that is the exponential decay, and live longer.
This means that it is reasonable to analyze conditions allowing for that.
We analyze properties of an ensemble of Universes with unstable vacua considered as an ensemble of  unstable systems
from the point of view of the quantum theory
of unstable states. We found some symmetry relations for quantities characterizing the metastable state. We also
found a relation linking the decay rate, ${\it\Gamma}$ of the metastable vacuum state with the Hubble parameter $H(t)$,
 which may help to explain why a universe with an unstable vacuum that lives longer then  the canonical decay times need not decay.

\end{abstract}
\noindent
Key words: {\em Unstable (false) vacuum, Quantum decay process, Cosmological constant problem}\\

\section{Introduction}

In cosmology a discussion of a problem of the false vacuum and the   possibility of its decay began  from the papers by Coleman and  his colleagues
\cite{Coleman,Callan,Col2}. Krauss in \cite{Krauss} analyzing properties of the false vacuum as a quantum unstable (quasi--stationary) state
 $|M\rangle$ drawn an attention to the problem that  there may exist universes in which the lowest energy state is  the false vacuum state and such that they
can survive up to the times much later than times $t$ when the canonical exponential decay law holds (see also \cite{Winitzki}).
 The study of cosmological models with unstable vacua  has became  particularly important in the context of
the discovery of the Higgs boson and of finding its mass $m_{H}$ \cite{Ade,Chat} to be $125.1 \pm 0.14$ [GeV] and the top quark mass $m_{t} \simeq 172.76 \pm 0.30$ [GeV] \cite{pdg-2020}. It is because the Standard Model calculations performed for the Higgs particle suggest that  the electroweak vacuum is unstable if the mass of the Higgs
particle is around  125 --- 126 GeV (see eg.  \cite{Deg,But,Isidori,bamba,graef,Spencer,Kob,Esp,Wei,Ema,Esp1}), which means that our Universe may be the Universe with unstable vacuum. For this reason  various mechanisms  slowing the vacuum decay down
 or even stopping it,
 have been discussed in many papers
 (see e. g.  \cite{Mar,Stojk2} and and also \cite{JK,De-Ch,Bur1,Bur2} and references therein).

In this paper we analyze multiverse made up of ensembles  of stable and unstable universes. The property of the universe "to be unstable" or "to be stable" is determined by properties of the vacuum state: If it is a false vacuum then it is unstable and decays into the true vacuum state and thus this universe decays too. The decay of the false vacuum is a quantum decay process and in this paper we will use this fact as an assumption.
Any quantum decay process, whether it is a decay of a particle or a decay of an excited level in an atom, or of the
decay of the metastable false vacuum, no matter how, (e. g. via
the quantum tunneling
through a potential barrier), must exhibit all the general properties resulting form the quantum theory of unstable states.
Therefore, in our opinion,
the quantum theory of unstable states seems to be an appropriate tool for the general analysis of the decay process of the false vacuum state and
 can help to understand and explain the various subtleties and properties of this process.
The vacuum decay plays an extremely important role in cosmology. It cannot be ruled out that without the decay of a metastable  vacuum it will be  impossible to explain some issues, as stated in \cite{Esp1} at the end of Sec. 6, where one can find the following sentence:
{\em
Assuming that the present
acceleration of the universe is due to a small cosmological constant, and accepting the conjecture that quantum gravity is ill--defined in a de Sitter space, we argue that vacuum decay is a necessary way out for the universe.}
Now, suppose, following the idea of Krauss and Dent \cite{Krauss}, that certain universes were lucky enough to survive the times of the canonical decay and they are still alive. (The {\em canonical decay times} means times
 when the decay law (the survival probability) has an exponential form to very good approximation).
 This idea can be applied to our Universe
if to assume that its current age is longer than  the canonical decay times of the false vacuum state.
It is worth noting here that
 there are  cosmological models under study in which the lifetime of a false vacuum is very short, and even significantly shorter
than the duration of the inflationary phase (see e.g. \cite{Branchina1,Branchina2}).
The important question is what conditions should be satisfied  in order that in such and similar cases
some universes could survive up to times much later than the canonical decay times and how long they are able to survive.
Here we attempt to clarify this issue
 considering unstable universes as an ensemble of unstable quantum particles and analyzing their behavior at very late times.
The tools we use for this purpose are the general properties of the quantum decay law, the decay rate, ${\it\Gamma}$, and the energy of the system in a metastable state in the region of very long times. From the general principles of the quantum theory it results that the decay rate depends on time, ${\it\Gamma} = {\it\Gamma}(t)$, and ${\Gamma}(t) \to 0$ as $t \to \infty$ whereas at canonical decay times ${\it\Gamma}(t) \simeq {\it\Gamma}^{0} = const.$ to very good approximation. In my opinion, these properties of the decay rate ${\it\Gamma}(t)$ may cause a universe whose vacuum is a false vacuum to survive longer than the lifetime of this false vacuum.

The paper is organized as follows: In Sec. 2 one can find a quantum description of the decay process and parameters characterizing this process.
In Sec. 3 a simplified toy model of the combined process of the expansion of a universe with unstable vacuum and of the quantum decay process of the unstable vacuum state is analyzed. Sec. 4 contains analysis of the long time properties of the survival amplitude and connected with  these properties a behavior of the decay rate as a function of time $t$. Properties of the energy of the metastable vacuum state as a function of time $t$ and the related properties of the density of the vacuum energy are  considered  in Sec. 5. Sec. 6 contains a discussion and conclusions.

\section{Preliminaries: Quantum description of the decay process}

From experiments it is known that for some unstable systems decay process is relatively fast or very fast while  for others it is slow or very slow.
The rate of this process is  characterized by  parameter called the "lifetime" or the "decay rate".
In decay experiments one has an ensemble of unstable physical systems in a certain area, which is surrounded by counters  that detect decay products. The counting rate, i.e. the number of decay per second $\frac{\delta N(t)}{\delta t}$, is proportional to the number of unstable particles $N(t)$ in a given volume at  instant $t$. The proportionality coefficient,
\begin{equation}
    \lambda =  \frac{\frac{\delta N(t)}{\delta  t}}{N(t)}, \label{gamma-1}
\end{equation}
is connected with the average lifetime (or simply lifetime) of the unstable objects considered (see, e.g.  \cite{bohm} for a discussion).
Indeed, if $N(t)$ is very large, then
the ratio of $N(t)$ by the initial number $N_{0}$ of such object at the initial instant $t_{0}^{init}$,
$N_{0}  = N(t_{0}^{init})$, in this area is the  probability, ${\cal P}(t)$, of finding an unstable object undecayed in this area at a given instant of time $t$ (i.e.,  the survival probability  ${\cal P}(t)$).  There is ${\cal P}(t) \simeq N(t) /N_{0}$ and  $\lim_{t\to\infty}\,N(t) =0$ so ${\cal P}(\infty) = 0$ and ${\cal P}(t_{0}^{init}) = 1$.
The number of decays $\delta N(t)$ per unit of time $\delta t$ equals: $\delta N(t) \simeq N(t ) - N(t+ \delta t)$. There is
$N(t) > N(t +\delta t)$ in the case of decay processes and thus
\begin{equation}
\lim_{\delta t \to 0}\,\frac{\delta N(t)}{\delta  t} = -\, \frac{d N(t)}{d  t}.
\end{equation}
The solution of Eq. (\ref{gamma-1}) in the case $\delta t \to 0$ takes the following form
\begin{equation}
\frac{N(t)}{N_{0}} = e^{\textstyle{- \lambda t}}.
\end{equation}
So, in this case there is ${\cal P}(t) \simeq \exp\,[- \lambda t]$
and the density of the probability of the decay at time $t$ during the time interval $t + dt$, $\rho_{\cal P}(t)$, equals $\rho_{\cal P}(t) = -\frac{d {\cal P}(t)}{dt} \equiv \lambda\,\exp[-\lambda t] \equiv \lambda\,{\cal P}(t)$. Taking for simplicity $t_{0}^{init} = 0$ it is easy to verify that
$\int_{0}^{\infty}\,\rho_{\cal P}(t)\,dt = 1$ as it should be. Using $\rho_{\cal P}(t)$ and keeping for a moment $t_{0}^{init} = 0$ one can find the average lifetime,
\begin{equation}
\tau = \langle t \rangle = \int_{0}^{\infty}\,t\,\rho_{\cal P}(t)(t)\,dt \equiv \frac{1}{\lambda}.
\end{equation}
Thus, in general
\begin{equation}
\lambda \equiv \frac{1}{\tau} = - \,\frac{\frac{d {\cal P}(t)}{dt}}{{\cal P}(t)} \stackrel{\rm def}{=} \frac{{\it\Gamma}}{\hbar},\label{gamma-2}
\end{equation}
where ${\it\Gamma}$ is the decay rate.

Within the quantum theory, similarly as in the case of the classical physics, the number of unstable particles $N(t)$, which at the time $t$ can be found in the area considered, is equal to the product
of the probability, ${\cal P}(t)$, of finding an unstable object undecayed in this area at a given instant of time $t$, (i. e. of the survival probability  ${\cal P}(t)$), and the initial number $N_{0}$ of such objects:
\begin{equation}
 N(t) = {\cal P}(t)\,N_{0}. \label{PN(0)}
 \end{equation}
where the survival probability ${\cal P}(t)$ (or the decay law)  is defined as
follows:
\begin{equation}
{\cal P}(t)  = |A(t)|^{2}. \label{P(t)} 
\end{equation}
and
\begin{equation}
A(t) = \langle M|M (t)\rangle,\label{a(t)}  
\end{equation}
is the survival amplitude,  $|M \rangle$ is the unstable (metastable) state under considerations,
$ |M\rangle \in \mathfrak{H}$ (where $\mathfrak{H}$ is the Hilbert space of states of the considered system), and
 $|M (t)\rangle$ is the solution of the Schr\"{o}dinger equation
\begin{equation}
i\,\hbar\,\frac{\partial }{\partial t}|M (t)\rangle = {\cal H} \, |M (t)\rangle, \label{Schrod} 
\end{equation}
for the initial condition $|M (t_{0}^{init}) \rangle = |M\rangle$. Here  ${\cal H}$ is the total self--adjoint Hamiltonian for the system under consideration and  $t_{0}^{init}$ is the initial instant. The vector  $|M (t)\rangle =  \exp\,[-\,\frac{i}{\hbar}(t-t_{0}^{init}){\cal H}]\,|M\rangle \equiv |M(t - t_{0}^{init})\rangle$ is the solution of Eq. (\ref{Schrod}).

It is easy to find that
\begin{equation}
A(t) \equiv A(t - t_{0}^{init})  = \langle M| \exp\,[-\,\frac{i}{\hbar}(t-t_{0}^{init}){\cal H}]\,|M\rangle  \equiv A^{\ast}[-(t - t_{0}^{init})]. \label{A(-t)}
\end{equation}
So, there are some symmetries of quantities characterizing the decaying state. The first one is given by Eq. (\ref{A(-t)}). The second one is a direct consequence of Eq (\ref{A(-t)}).  Namely using (\ref{A(-t)}) one finds that there is the following symmetry,
\begin{equation}
{\cal P}(t) \equiv A(t)\,A^{\ast}(t) = A(t) \,A(-t) = {\cal P}(-t). \label{P(-t)}
\end{equation}

From (\ref{gamma-2}), (\ref{P(t)})
we obtain that,
\begin{equation}
\frac{{\it\Gamma}}{\hbar} \equiv \frac{{\it\Gamma}_{M}(t)}{\hbar} = - \Big( \frac{1}{A(t)}\,\frac{\partial A(t)}{\partial t} + \frac{1}{A^{\ast}(t)}\,\frac{\partial A^{\ast}(t)}{\partial t}\Big).
\label{gamma-3}
\end{equation}

Using (\ref{A(-t)}) and (\ref{gamma-3}) another symmetry is easy to find. This time   for ${\it\Gamma}_{M}(t)$. There is
\begin{equation}
{\it\Gamma}_{M}(t)  = \,-\,{\it\Gamma}_{M}(-t). \label{G(-t)}
\end{equation}

If to define the following quantity \cite{pra}:
\begin{equation}
h_{M}(t) = \frac{i\hbar}{A(t)}\,\frac{\partial A(t)}{\partial t}
\label{h}
\end{equation}
then the relation (\ref{gamma-3})
means that simply
\begin{equation}
{\it\Gamma}_{M}(t) = - 2 \Im\,[h_{M}(t)], \label{gamma-h}
\end{equation}
where $\Im\,[z]$  denotes the imaginary parts of $z$ (similarly, $\Re\,[z]$ is the real part of $z$).

Note that  one can find also the symmetry for $h_{M}(t)$ that results directly from Eq  (\ref{A(-t)}) and from the definition (\ref{h}) of $h_{M}(t)$. There is
\begin{equation}
h_{M}^{\ast}(t) = h_{M}( -t). \label{h(-t)}
\end{equation}

From basic principles of the quantum theory it follows that the
amplitude $A(t)$, and thus the decay law ${\cal P}(t)$ of the
unstable state $|M\rangle$, can be completely determined by the
density of the energy distribution $\omega({ E})$ for the system
in this state \cite{Fock,Fonda}
\begin{equation}
A(t) = \int_{Spec.({\cal H})} \omega({ E})\;
e^{\textstyle{-\,\frac{i}{\hbar}\,{ E}\,(t-t_{0}^{init})}}\,d{ E} \equiv A(t - t_{0}^{init}).
\label{a-spec}
\end{equation}
where $\omega({\cal E}) \geq 0$. 

In \cite{Khalfin}
assuming that the spectrum of ${\cal H}$ must be bounded
from below,
$Spec.({\cal H}) \stackrel{\rm def}{=} \sigma({\cal H})= [E_{min},\infty)$ and $E_{min} > -\infty$,
and using the Paley--Wiener Theorem  \cite{Paley}
it was proved that in the case of unstable
states there must be
\begin{equation}
|A(t)| \; \geq \; B\,e^{\textstyle - b \,t^{q}} \;\;\;\;
{\rm  for} \;\;\;\; |t| \rightarrow \infty, \label{a-kh}
\end{equation}
(where $B > 0,\,b> 0$ and $ 0 < q < 1$).
This means that the decay law ${\cal P}(t)$ of unstable
states decaying in the vacuum can not be described by
an exponential function of time $t$ if time $t$ is suitably long, $t
\rightarrow \infty$, and that for these lengths of time ${\cal
P}(t)$ tends to zero as $t \rightarrow \infty$  more slowly
than any exponential function of $t$.
The analysis of the models of
the decay processes shows that
${\cal P}(t) \simeq
\exp[- {\it\Gamma}_{M}^{0} t/\hbar]$,
to a very high accuracy
from $t$ suitably later than the initial instant $t_{0}^{init}$
up to
$ t \gg \tau_{M}$, (where $\tau_{M} = \hbar/{\it\Gamma}_{M}^{0}$ is the life--time of the state $|M\rangle$, ${\it\Gamma}_{M}^{0}$ is the decay width of the unstable state $|M\rangle$ calculated within the one pole approximation \cite{WW}),
and smaller than $t = T_{1}$, where $T_{1}$  denotes the
time $t$ at which the non--exponential deviations of $A(t)$ begin to dominate.

In general,
in the case of quasi--stationary (metastable) states it is convenient to express $A(t)$ in the
following form
\begin{equation}
A(t) = A_{c}(t) + A_{lt}(t), \;\; ({\rm for}\;\;t\gg \tau_{M}), \label{a-exp+}
\end{equation}
where $A_{c}(t)$ is the exponential part of $A(t)$, that is
$A_{c}(t) =
N\,\exp[-\,\frac{i}{\hbar}(t-t_{0}^{init})(E_{M}^{0}  -  \frac{i}{2}\,{\it\Gamma}_{M}^{0})] $,
($N$ is the normalization constant, $E_{M}^{0}$ is the energy of the system in the unstable state $|M\rangle$ calculated within the one pole approximation),
 and $A_{lt}(t)$ is the
late time non--exponential part of $A(t)$.
For times $t \sim \tau_{M}$:
$|A_{c}(t)| \gg |A_{lt}(t)|$. Using (\ref{a-exp+}) one finds that
\begin{equation}
{\cal P}(t) = |A(t)^{2} = |A_{c}(t)|^{2} + 2 \Re\,[A_{c}(t)\,A_{lt}^{\ast}(t)] + |A_{lt}(t)|^{2}. \label{a-c+a-lt}
\end{equation}
The solution, $t$,  of the equation
\begin{equation}
|A_{c}(t)|^{2} = 2 \Re\,[A_{c}(t)\,A_{lt}^{\ast}(t)], \label{T1}
\end{equation}
(let us denote it as $t=T_{1}$) is usually considered as an approximate, conventional end of the canonical phase of a decay process, where the survival probability has an exponential form:  For $t < T_{1}$ there is
${\cal P}(t) \simeq |A(t)|^{2} = \exp\,[-i \frac{t}{\hbar} {\it\Gamma}^{0}_{M}]$ to a very good approximation. Solving the following equation
\begin{equation}
 2 \Re\,[A_{c}(t)\,A_{lt}^{\ast}(t)] =  |A_{lt}(t)|^{2}, \label{T2}
\end{equation}
one finds the time $t=T_{2}$. The time $T_{2}$   is the time from which the late time phase of the decay process begins: For $t > T_{2}$  the survival probability has a form of powers of $(1/t)$. The transition phase of a decay process is the epoch when time $t$ is passing the time--interval $  (T_{1}, T_{2})$. At this point, a mention should also be made of the so-called "cross--over time" used by some author (see, e. g. \cite{rothe}).
The crossover time, denoted above as $T$, is the time  when contributions to the survival probability ${\cal P}(t)$ of
its exponential (canonical) and late time non-exponential parts are the same:
\begin{equation}
|A_{c}(t)|^{\,2} = |A_{lt}(t)|^{\,2}, \label{A-exp=A-lt}
\end{equation}
and $T$ is the solution of this equation. There is  $T_{1} < T < T_{2}$.

At this point, it should be noted that the consideration of asymptotic late time properties of the amplitude $A(t)$ and the quantities defined within the use of  $A(t)$ is justified by experimental results. Namely,
in an experiment described  in the Rothe paper
\cite{rothe},
the experimental evidence of deviations from
the exponential decay law at long times,
much later than the crossover time $T$,
was reported.

From relations (\ref{P(t)}), (\ref{gamma-3}), (\ref{gamma-h}) it is seen that
the amplitude $A(t)$ contains information about
the decay law ${\cal P}(t)$ of the state $|M\rangle$
and about the decay rate $\it\Gamma (t)$.
It was also shown that using (\ref{h}) the information  about the energy $E_{M}(t)$ of the system in the unstable state considered can also  be extracted from the survival amplitude $A(t)$:
The energy of the system in the unstable state $|M\rangle$ (the instantaneous energy), $E_{M}(t)$, is equal to the real part of the effective hamiltonian $h_{M}(t)$ (see eg. \cite{pra}),
\begin{equation}
E_{M}(t) = \Re \, [h_{M}(t)]. \label{E=Re-h}
\end{equation}
and in general we have.
\begin{equation}
h_{M}(t) = E_{M}(t) - \frac{i}{2}\,{\it\Gamma}_{M}(t). \label{h=E-iG}
\end{equation}
There is the following symmetry for $E_{M}(t)$ completing the symmetry relation (\ref{G(-t)}), which results directly form Eqs (\ref{h}), (\ref{A(-t)}):
\begin{equation}
E_{M}(t) = E_{M}(-t). \label{E(-t)}
\end{equation}

Now let us focus an attention on the survival amplitude $A(t)$ and on the survival probability ${\cal P}(t)$ given by (\ref{a(t)}) and (\ref{P(t)})
 and on the description of the decay of a metastable false vacuum.
${\cal P}(t)$ is the probability to find the system at time $t$ in the metastable state $|M (t_{0}^{init})\rangle \equiv |M \rangle$  prepared in the initial instant $t_{0}^{init}  >0$. If there were a suitable large number $N_{0}$ of identical unstable  objects at the initial instant $t_{0}^{init}$ then according to (\ref{PN(0)}) one should detect  $N(t) = {\cal P}(t)\,N_{0} \,<\,N_{0}$  of them at $t > t_{0}^{init}$ . There is no such a simple correspondence of ${\cal P}(t)$ with the results of measurements when
one is able to prepare only
one particle (or a few particles) at $t_{0}^{init}$. On the other hand if one is able to prepare at $t_{0}^{init}$ a system containing only one unstable object and  to make a large number $N_{0}$ of indistinguishable copies of this system
then the problem reduces to the previous one: $N(t) = {\cal P}(t)\,N_{0}$ copies of the system will contain this unstable object undecayed at $t > t_{0}^{init}$.  When there are no $N_{0} \gg 1$ copies of the system at $t =t_{0}^{init}$ but one has to deal with only one particle system then one can never be sure whether one will detect this particle undecayed at $t > t_{0}^{init}$ or not. This same concerns a universe with the metastable (false) vacuum:
One can expect that  an ensemble of $N_{0}$ universes with unstable vacua will  behave analogously as a system containing $N_{0}$ unstable objects.
So, let $|M\rangle \equiv |0^{M}\rangle$ be the metastable  (false) vacuum state of a universe considered and $|0^{M}\rangle \neq |0^{\;true}\rangle$, (where $|0^{\;true}\rangle$ is the true ground state describing the state in which the energy of the system under considerations has the  absolute minimum). Let us assume that this universe  was created at the instant $t = t_{0}^{init} > 0$ and the volume occupied by this universe  at $t= t_{0}^{init}$ was  ${V_{0}^{init} = V(t)\,}\vline_{\,t=t_{0}^{init}}$. Thus in fact one should take into account that there is  $|0^{M}\rangle \equiv |0^{M};V_{0}^{init}\rangle$, where $|0^{M};V_{0}^{init}\rangle$ is the vacuum state of the universe of the volume $V_{0}^{init}$. It is convenient to choose the normalization condition for $|0^{M};V_{0}^{init}\rangle$ in the following form,
\begin{equation}
\langle V_{0}^{init};\,^{M}0|0^{M};V_{0}^{init}\rangle = 1. \label{<|>=1}
\end{equation}
In this case
an analysis of the survival probability ${\cal P}(t)$
can not give a conclusive answer whether the universe of the volume $V_{0}^{init}$
will still exist in the state $|0^{M};V_{0}^{init}\rangle$ at instant $t > t_{0}^{init}$ or not.
The problem becomes much more complicated if to take into account that in addition to
the pure quantum tunneling process leading to decay of the false vacuum state \cite{Coleman,Callan,Col2}  there exists another completely different process forcing the universe of the volume $V_{0}^{init}$ to expand.
This effect was considered in \cite{Krauss}, where
Krauss and Dent analyzing a false vacuum decay
pointed out that in eternal inflation, even though regions of false vacua by assumption
should  decay exponentially, gravitational effects force the space region of the volume $V_{0}^{init}$ that has not decayed yet
to grow exponentially fast.
This effect causes that many false vacuum regions or many universes forming a multiverse
can survive up to the times much later than times when the exponential decay law
holds.
What is more,
the particle physics can provide us with hints suggesting what may happen in such or similar cases:
A free neutron is unstable and decays but the neutron inside a nucleus is subjected
to other additional interactions and does not decay.
These processes both can be described using the survival amplitude (\ref{a(t)}), (\ref{A(-t)})
with suitable Hamiltonians ${\cal H}$. There is ${\cal H} = {\cal H}_{W}$ in the case of the free neutron  and
there is ${\cal H}= {\cal H}_{W} + {\cal H}_{S}$ for the neutron inside a nucleus. Here ${\cal H}_{W}$ describes weak interactions where as ${\cal H}_{S}$ denotes strong and electromagnetic interactions. For the free neutron we have $|A(t)|^{2}\;\to\;0$ as $t \,\to\,\infty$. This property is not the case of the neutron inside the nucleus. In general, when an unstable particle is  subjected to different interactions described  by suitable commuting Hamiltonians, then it may happen that the decay   process can be slowed or even stopped.
Similarly, as it was shown in \cite{Col2}, the gravitation my stop the decay of the false vacuum.
So when analyzing   the stability of the false vacuum state by means of the survival amplitude $A(t)$ the correct conclusion can not be drawn if to use  only the Hamiltonian ${\cal H}$ describing the "pure" decay through quantum tunneling. One can expect that the correct result can be obtained if to replace this ${\cal H}$ in (\ref{a(t)}), (\ref{A(-t)}) by the summ ${\cal H} + {\cal H}_{V}$, where ${\cal H}_{V}$ describes more or less accurately the expansion process of the volume $V_{0}^{init}$.

\section{A simplified toy model}

Astrophysical observations lead to the conclusions that our Universe is expanding in time. In \cite{Krauss} an observation was made that in inflationary processes, even if some space regions of false (unstable) vacua decay exponentially, gravitational effects force space in a region that did not have time to decay, to grow exponentially fast (see also \cite{Winitzki}). So, in general the expansion process affect the process of decay of the universes (domains) with the false vacua.
The problem is how to describe this expansion so that variations in time of the volume $V(t)$ occupied by the Universe had the form of Schr\"{o}dinger Equation (\ref{Schrod}) or a similar form with a suitable effective hamiltonian ${\cal H}_{V}$. The volume $V(t)$ is an increasing function of time $t$ in the present epoch, so its evolution is non--unitary and ${\cal H}_{V}$ can not be hermitian. The non--unitary evolution operator solving the  Schr\"{o}odinger--like equation with this ${\cal H}_{V}$ and  acting on the initial state $|0^{M};V_{0}^{init}\rangle$ should transform this state into the vector $|\psi (t)\rangle = |0^{\,false}(t); V(t)\rangle \equiv \alpha\;[V(t)]^{1/2}\,\exp\,[-\,\frac{i}{\hbar}(t- t_{0}^{init}){\cal H}]|0^{\,false};V_{0}^{init}\rangle$, where $\alpha$ is a complex or real number.
The simplest ${\cal H}_{V}$, which seems to be sufficient for the simplified qualitative analysis of the problem, may be chosen as follows,
\begin{eqnarray}
{\cal H}_{V} \equiv {\cal H}_{V}(t) &=& (E_{V}\, +\, i\,\hbar\,\frac{d}{dt}\ln\,[ a^{3/2}(t)])\,\mathbb{I} \label{Hv} \\
&=& (E_{V}\, +\, i\,\hbar\,\frac{3}{2}\,H(t))\,\mathbb{I}, \label{Hv1}
\end{eqnarray}
where  $a(t) = R(t)/R_{0}$ is the scale factor, $R(t)$ is the proper distance at
epoch $t$, $R_{0}= R(t_{0})$ is the distance at the reference time $t_{0}$,
(it can be also interpreted as the radius of the Universe now)  and here $t_{0}$ denotes  the present epoch (see, e. g. \cite{Cheng}),
 $H(t) = \frac{\dot{a}(t)}{a(t)}$ is the Hubble parameter, $\dot{a}(t) = \frac{d}{dt}a(t)$ (in the general case $\dot{f}(t) \equiv \frac{df(t)}{dt}$),
$\mathbb{I}$ is the unit operator, ${\cal H}_{V}$ is the non--hermitian effective Hamiltonian, $E_{V}$ is a real parameter having dimension of the energy.
The scale factor $a(t)$ is a solution of
Einstein's equations, which
 with the Robertson---Walker metric
in the standard form of Friedmann equations \cite{Cheng,Sahni},
look as follows:
The first one,
\begin{equation}
\frac{{\dot{a}}^{2}(t)}{a^{2}(t)}  + \frac{kc^{2}}{R_{0}^{2}\,a^{2}(t)} = \frac{8\pi G_{N}}{3}\,\rho +\frac{\Lambda\,c^{2}}{3}, \label{Fr1}
\end{equation}
and the second one,
\begin{equation}
\frac{{\ddot{a}}(t)}{a(t)} =\,-\,\frac{4\pi G_{N}}{3}\,\left(\frac{3p}{c^{2}} +  \rho \right) + \frac{\Lambda\,c^{2}}{3}. \label{Fr2}
\end{equation}
where the parameter $\Lambda$ is known as the cosmological constant,
$\rho$ and $p$ are mass density  and pressure respectively, $k$ denote the curvature signature,
 The pressure $p$ and the density $\rho$ are
related to each other through the equation of state, $p = w\rho \,c^{2}$, where $w$ is constant \cite{Cheng}. There is $w=0$ for a dust (for a matter dominated era), $w=1/3$ for a radiation and $w = -1$ for a
vacuum energy.

The volume $V(t)$ equals: $V(t) = \frac{4}{3}\pi [R(t)]^{3} \equiv \frac{4}{3}\pi [a(t)R_{0}]^{3}$, and similarly, $V_{0}^{init} = \frac{4}{3}\pi [a(t_{0}^{init})R_{0}]^{3}$. Therefore
\begin{equation}
V(t) = \left[\frac{a(t)}{a(t_{0}^{init})}\right]^{3}\,V_{0}^{init}. \label{V(t)}
\end{equation}
We are looking for the solutions of the Schr\"{o}dinger equation with the Hamiltonian $({\cal H} +{\cal H}_{V})$ and a matrix element of the form $\;\langle V_{0}^{init};^{false\,}0|\psi (t)\rangle$ with $|\psi (t_{0}^{init})\rangle = |0^{M};V_{0}^{init}\rangle$. So we need solutions of the following equation
\begin{equation}
i\hbar \frac{d}{dt}\,|\psi (t)\rangle = ({\cal H} + {\cal H}_{V})\,|\psi (t)\rangle, \label{d-psi}
\end{equation}
with the initial condition $|\psi (t_{0}^{init}) \rangle = |0^{M};V_{0}^{init}\rangle$. Here ${\cal H}$ is a hermitian operator (Hamiltonian) responsible for the decay of the false vacuum state $|0^{M};V_{0}^{init}\rangle$ and $[{\cal H},{\cal H}_{V}] = 0$. Now, let $|\psi (t)\rangle$ be of the form
\begin{equation}
|\psi (t)\rangle = e^{\textstyle{-\,\frac{i}{\hbar}(t-t_{0}^{init}){\cal H}}}\,|M(t)\rangle, \label{psi}
\end{equation}
and
\begin{equation}
|M (t_{0}^{init})\rangle = |0^{M};V_{0}^{init}\rangle. \label{M(0)}
\end{equation}
Inserting (\ref{psi}) into (\ref{d-psi}) one obtains
\begin{eqnarray}
{\cal H}\,    e^{\textstyle{-\,\frac{i}{\hbar}(t-t_{0}^{init}){\cal H}}}\,|M(t)\rangle &+& e^{\textstyle{-\,\frac{i}{\hbar}(t-t_{0}^{init}){\cal H}}}\,i\hbar \frac{d}{dt}\,|M(t)\rangle  = \nonumber \\ &=& {\cal H}\, e^{\textstyle{-\,\frac{i}{\hbar}(t-t_{0}^{init}){\cal H}}}\,|M(t)\rangle \nonumber \\
&& +\, e^{\textstyle{-\,\frac{i}{\hbar}(t-t_{0}^{init}){\cal H}}}\,{\cal H}_{V}\,|M(t)\rangle,
 \end{eqnarray}
This means that our problem reduces to  finding a solution of the following equation
\begin{equation}
i\hbar \frac{d}{dt}\,|M(t)\rangle = (E_{V}\, +\, i\hbar\,\frac{d}{dt}\ln\,[a^{3/2}(t)])|M(t)\rangle. \label{M(t)}
\end{equation}
Putting
\begin{equation}
|M(t) \rangle = f(t)\,|M(t_{0}^{init})\rangle \equiv f(t)|0^{M};V_{0}^{init}\rangle, \label{M(t)=f}
\end{equation}
where $f(t)$ is a real or complex  scalar function and $f(t_{0}^{init}) = 1$,
we can
rewrite Eq. (\ref{M(t)}) as follows
\begin{equation}
i\hbar\frac{df(t)}{dt}\,|0^{M};V_{0}^{init}\rangle =  (E_{V}\, +\, i\hbar\,\frac{d}{dt}\ln\,[a^{3/2}(t)])\,f(t)\,|0^{M};V_{0}^{init}\rangle. \label{df}
\end{equation}
A solution, $f(t)$,  of this equation
is
\begin{eqnarray}
f(t) &=& N_{f}\,e^{\textstyle{-\,\frac{i}{\hbar}(t-t_{0}^{init})E_{V}}}\,e^{\textstyle{+ \int_{t_{0}^{init}}^{t}\,\frac{d}{dx}\ln\,[a^{3/2}(x)]\, dx}}\,f(t_{0}^{init})
\nonumber \\
&\equiv& N_{f}\,
e^{\textstyle{-\,\frac{i}{\hbar}(t-t_{0}^{init})E_{V}}}\,\left[\frac{a(t)}{a(t_{0}^{init})}\right]^{3/2}, \label{f}
\end{eqnarray}
where $N_{f}$ is a normalization factor.
Now inserting this $f(t)$ into (\ref{M(t)=f}) and then using (\ref{psi}) we obtain the solution, $|\psi (t)\rangle$, of Eq. (\ref{d-psi}),
\begin{equation}
|\psi (t) \rangle = N_{f}\, e^{\textstyle{-\,\frac{i}{\hbar}(t-t_{0}^{init})E_{V}}}\,\,\left[\frac{a(t)}{a(t_{0}^{init})}\right]^{3/2}\,e^{\textstyle{-\,\frac{i}{\hbar}(t- t_{0}^{init}){\cal H}}}\,|0^{M};V_{0}^{init}\rangle. \label{psi(t)}
\end{equation}
Thus
\begin{eqnarray}
\langle V_{0}^{init};\,^{M}0|\psi (t)\rangle &\equiv &N_{f} e^{\textstyle{-\,\frac{i}{\hbar}(t-t_{0}^{init})E_{V}}}\,\left[\frac{a(t)}{a(t_{0}^{init})}\right]^{3/2} \times \nonumber \\ && \times \langle V_{0}^{init};\,^{M}0|\,e^{\textstyle{-\,\frac{i}{\hbar}(t-t_{0}^{init}){\cal H}}}\,|0^{M};V_{0}^{init}\rangle,
\label{0|psi(t)}
\end{eqnarray}
 and
 \begin{equation}
 \Pi(t) \stackrel{\rm def}{=} |\;\langle V_{0}^{init};\,^{M}0|\psi (t)\rangle\,|^{2} \,\equiv N_{f}^{2}\,\,\left[\frac{a(t)}{a(t_{0}^{init})}\right]^{3}\,{\cal P}_{0}(t). \label{pi}
 \end{equation}
Here
 \begin{equation}
{\cal P}_{0}(t) \stackrel{\rm def}{=}  |\;\langle V_{0}^{init};\,^{M}0|\,e^{\textstyle{-\,\frac{i}{\hbar} (t-t_{0}^{init}) {\cal H}}}\,|0^{M};V_{0}^{init}\rangle\,|^{2}, \label{P_0}
 \end{equation}
 is the survival probability of the system in the initial false vacuum state $|0^{M};V_{0}^{init}\rangle$ assuming that volume $V_{0}^{init}$ occupied by  this system remains unchanged. The function $\Pi (t)$ describes the combined effect of the processes of a decay and an expansion of the initially created universes.

 There is $V(t) = \left[\frac{a(t)}{a(t_{0}^{init})}\right]^{3}\,V_{0}^{init}$ but the use of normalization factor, $N_{f}$, allows  us to write  volume $V(t)$ as $V(t) \equiv \left[\frac{a(t)}{a(t_{0}^{init})}\right]^{3}$.  So
 \begin{equation}
 \Pi (t)  \equiv N_{f}^{2}\, V(t)\,{\cal P}_{0}(t), \label{pi(t)}
 \end{equation}
and
\begin{equation}
\Pi (t)\, N_{0}\, \equiv N_{f}^{2}\,\mathbb{V}(t)\,{\cal P}_{0}(t), \label{piN0}
\end{equation}
where $N_{0}$ is the number of universes of volume $V_{0}^{init}$ created at the initial instant $t_{0}^{init}$ with the vacua described by
$|0^{M};V_{0}^{init}\rangle$ and $\mathbb{V}(t)$ is volume occupied by all these universes  at the instant $t > t_{0}^{init}$ and it corresponds with $N(t)$ in (\ref{PN(0)}).
Relations (\ref{pi(t)}), (\ref{piN0}) describe in our simplified toy model the combined effect of the processes of a decay and of an expansion of the initially created universes  of volumes $V_{0}^{init}$. In the case when the decay process is the dominant process then $\Pi (t)$  appearing in (\ref{piN0}) is a decreasing function of time $t$  and it tends to zero as $t \to \infty$. If the expansion process prevails over the decay process or these processes both are in balance then $\Pi (t)$ is non--decreasing function of $t$. In such a case
\begin{equation}
\frac{d}{dt}\Pi (t) \geq 0,
\end{equation}
that is,
\begin{eqnarray}
\frac{d}{dt}[\Pi (t) N_{0}] &\equiv& N_{f}^{2}\,\Big(\frac{\dot{\mathbb{V}}(t)}{\mathbb{V}(t)}\,+\,\frac{\dot{\cal P}_{0}(t)}{{\cal P}_{0}(t)}\Big)\,{\mathbb{V}}(t)\,{\cal P}_{0}(t) \nonumber \\
&=& N_{f}^{2}\,\Big(3\,\frac{\dot{a}(t)}{a(t)}\,-\,\frac{{\it\Gamma}_{M}(t)}{\hbar}\Big)\,{\mathbb{V}}(t)\,{\cal P}_{0}(t) \nonumber \\
&=& N_{f}^{2}\, \big(3\, H(t)\,-\,\frac{{\it\Gamma}_{M}(t)}{\hbar}\big)\,{\mathbb{V}}(t)\,{\cal P}_{0}(t)\;\geq\;0.
\end{eqnarray}
So, if there exists such time, say $t = T_{L} > 0$, that for all $t \geq T_{L}$ the relation,
\begin{equation}
d_{H,\it\Gamma} \stackrel{\rm def}{=}3\,H(t)\,-\,\frac{{\it\Gamma}_{M}(t)}{\hbar}\;\geq\;0, \label{main}
\end{equation}
is satisfied then the function $\Pi (t)$  is a non--decreasing function of time $t$ (it increases or is constant in time). This means that in such a case the decay process of the volumes $V(t) = \left[\frac{a(t)}{a(t_{0}^{init})}\right]^{3}\,V_{0}^{init}$ should be stopped.
Therefore if some universes   had the luck to survive until time  $T_{L}$ such that for all $t \geq T_{L}$ the relation (\ref{main}) is fulfilled then  later, when  $t >T_{L}$, these universes should be found undecayed.

\section{Late time properties of the decay rate ${\it\Gamma}_{M} (t)$ and related quantities}

As it was mentioned in Sec. 2,
the experimental evidence of deviations from
the exponential decay law at long times,
much later than the crossover time $T$,
was reported in
\cite{rothe}.
This result gives rise to problem which
is important for our considerations:
If (and how) deviations from the
exponential decay law at long times affect
the decay rate of the unstable state and the energy of the system in this state.

From the condition  (\ref{a-kh}) for the amplitude $A(t)$ and from (\ref{P(t)}) it results that at the long time region the lowest bound for the survival probability ${\cal P}(t)$ has the form
\begin{equation}
{\cal P}(t) \sim B^{2}\,e^{\textstyle{-2bt^{q}}}\;\;\;\;{\rm for}\;\;\;\;|t| \to \infty. \label{P(t)-as}
\end{equation}
This and the relation (\ref{gamma-2}) lead to the conclusion that (see \cite{pra})
\begin{equation}
{\it\Gamma}_{M} (t) \sim 2\hbar bq\,t^{q-1}\;\;\;\;{\rm for}\;\;\;\;|t| \to \infty, \label{G(t)-as}
\end{equation}
and thus ${\it\Gamma}_{M} (t) \to 0$ as $t \to \infty$ because $q < 1$. A more accurate estimation of ${\it\Gamma}_{M}(t)$ can be found using the amplitude $A(t)$ instead of the condition
(\ref{a-kh}) for the modulus $|A(t)|$ of $A(t)$.

So let us assume that we know the amplitude $A(t)$.  Equivalently it is sufficient to know  the energy distribution $\omega({ E})$ of the system in the unstable state considered: In such a case $A(t)$ can be calculated using (\ref{a-spec}). Then starting
with the $A(t)$ and using the expression (\ref{h}) one can
calculate the effective Hamiltonian $h_{M}(t)$ in a general case
for every $t$. Thus, one can find
the instantaneous
energy, $E_{M}(t)$, and the instantaneous decay rate, ${\it\Gamma}_{M}(t)$,
of the system in the metastable  state $|M\rangle$
for  canonical decay times, when  $ t \sim \tau_{M} <  T_{1}$,  for transition times $t \in (T_{1}, T_{2})$ and for asymptotically late times $ t  > T_{2}$
(for details see: \cite{urbanowski-2-2009,urbanowski-1-2009,giraldi}).

The integral representation (\ref{a-spec}) of $A(t)$ means that $A(t)$ is
the Fourier transform of the energy distribution function $\omega(E)$. Using this fact we can find
asymptotic form of $A(t)$ for $t \rightarrow \infty$, that is $A_{lt}(t)$ (see \cite{urbanowski-1-2009} for details):
As it  has been shown in \cite{urbanowski-1-2009}, if to assume that
$\lim_{ {E} \rightarrow { E}_{min}+}
\;\omega ({ E})\stackrel{\rm def}{=} \omega_{0}>0$ and
$\omega (E < E_{min}) = 0$
and
derivatives  $\omega^{(k)}({ E})$, ($k= 0,1,2, \ldots, n$),
are
continuous
in
$[ { E}_{min}, \infty)$, (that is
if
for ${ E} > { E}_{min}$ all
$\omega^{(k)}({ E})$
are
continuous
and all the limits 
$\lim_{{E} \rightarrow { E}_{min}+}\,\omega^{(k)}({E})$ exist),  and also
all these $\omega^{(k)}({ E})$
are
absolutely integrable functions, then
\begin{eqnarray}
A(t) \equiv A(t - t_{0}^{init}) \; & \begin{array}{c}
          {} \\
          \sim \\
          \scriptstyle{t \rightarrow \infty}
        \end{array} &
        \;- \frac{i\hbar}{t-t_{0}^{init}}\;e^{\textstyle{ -\,\frac{i}{\hbar} { E}_{min} (t-t_{0}^{init})}}\; \times \nonumber \\
       && \times  \sum_{k = 0}^{n-1}(-1)^{k} \,\big(\frac{i\hbar}{t-t_{0}^{init}}\big)^{k}\,\omega^{(k)}_{0}
        = A_{lt}(t),
        \label{a-omega}
\end{eqnarray}
where $\omega^{(k)}_{0}  \stackrel{\rm def}{=} \lim_{{E}\rightarrow {E}_{min}+}
\;\omega^{(k)} ({E})$ (see \cite{urbanowski-1-2009,giraldi}).

Bearing in mind the purpose of our considerations, which is to look from the point of view of the quantum theory of unstable states at the fate of the universe at times $t$ very distant from the moment of its creation, $t_{0}^{init}$,  we assume that $t > T_{2} \gg t_{0}^{init}$.
As a result we can write that $(t - t_{0}^{init}) \simeq t$ and  we will use  this conclusion in our late time asymptotic formulae for $ t \to \infty$ considered in this paper.

In the case of a universal more general form of $\omega (E)$, when
\begin{equation}
\omega ({ E}) = ( { E} - { E}_{min})^{\lambda}\;\eta ({ E})\; \in \; L_{1}(-\infty, \infty),
\label{omega-eta}
\end{equation}
where $0 < \lambda < 1$, and it is assumed that $\eta (E_{min}) > 0$, $\eta (E < E_{min}) = 0$ and derivatives $\eta^{(k)}({ E})$,
($k= 0,1,\ldots, n$),  
exist and they are continuous
in $[{E}_{min}, \infty)$, and  limits
$\lim_{{ E} \rightarrow {E}_{min}+}\;\eta^{(k)}({ E})$ exist,
$\lim_{{ E} \rightarrow \infty}\;( { E} - { E}_{min})^{\lambda}\,\eta^{(k)}({ E}) = 0$
for all above mentioned $k$,
there is,
\begin{eqnarray}
A(t) & \begin{array}{c}
          {} \\
          \sim \\
          \scriptstyle{t \rightarrow \infty}
        \end{array} &
        (-1)\,e^{\textstyle{- \frac{i}{\hbar}{ E}_{min} t}}\;
        \Big[
        \Big(- \frac{i \hbar}{t}\Big)^{\lambda + 1}  
         \Gamma(\lambda + 1)\;\eta_{0}\; \label{a-eta} \\
        && +\;\lambda\,\Big(- \frac{i \hbar}{t}\Big)^{\lambda + 2}  
        \; \Gamma(\lambda + 2)\;\eta_{0}^{(1)}\;+\;\ldots
        \Big]
         = A_{lt}(t),
            \nonumber
\end{eqnarray}
as it has been shown in \cite{urbanowski-1-2009}. Here  $\Gamma (z)$ is the Euler's Gamma Function.

Starting from the  asymptotic expression (\ref{a-eta}) for $A(t)$ and using (\ref{h})
after some algebra one finds  that in general for  times $t > T_{2}$  (see \cite{urbanowski-1-2009})
\begin{equation}
{h_{M}(t)\vline}_{\,t \rightarrow \infty} \simeq { E}_{min} + (-\,\frac{i\hbar}{t})\,c_{1} \,
+\,(-\,\frac{i\hbar}{t})^{2}\,c_{2} \,+\,\ldots, \label{h-infty-gen}
\end{equation}
where $ c_{i} = c_{i}^{\ast},\;\;i = 1,2,\ldots$ , (coefficients $c_{i}$ are determined by  $\omega (E)$).

This  last relation means that (see \cite{PLB-2014}),
\begin{eqnarray}
{\it\Gamma}_{M} (t) &\simeq & 2\,c_{1}\,\frac{\hbar}{t} \,-\,2c_{3} \,\frac{\hbar^{3}}{t^{3}}\ldots, \;\;\;({\rm for}
\;\;t > T_{2}), \label{G(t)-as}
\end{eqnarray}
and similarly
\begin{eqnarray}
{ E}_{M}(t) &\simeq & E_{min} \, - \,c_{2} \,\frac{\hbar^{2}}{t^{2}}\,+\,c_{4}\, \frac{\hbar^{4}}{t^{4}}\,+\,\ldots, \;\;\;({\rm for}
\;\;t >  T_{2}), \label{E(t)-as}
\end{eqnarray}
These properties take place for  all unstable states which survived up to times $t > T_{2}$.
From (\ref{E(t)-as}) it follows that
$\lim_{t \rightarrow \infty}\, {E}_{M}(t) = E_{min}$.

Note that the symmetry relations (\ref{G(-t)}), (\ref{h(-t)}) oraz (\ref{E(-t)})   also hold for the asymptotic expansions (\ref{h-infty-gen}), (\ref{G(t)-as}) and (\ref{E(t)-as}).

For the most general form (\ref{omega-eta}) of the density $\omega (E)$ (i. e. for $A(t)$ having the asymptotic
form given by (\ref{a-eta}) )  we have (see \cite{PLB-2014} and references herein).
\begin{equation}
c_{1} = \lambda + 1 > 0, \;\;\;\;c_{2} = (\lambda + 1)\,\frac{\eta^{(1)}(E_{min})}{\eta (E_{min})}\;>\; 0. \label{c-i}
\end{equation}
As an example let us consider a typical form of $\omega(E)$. Namely,
properties of metastable systems are described in mamy papers with sufficient accuracy using $\omega (E)$ having the form of
the  Breit--Wigner   energy distribution function, $\omega_{BW} (E)$,
\begin{equation}
\omega ({E}) \equiv \omega_{BW} (E) =
  \frac{N}{2\pi}\,  {\it\Theta} ( E - E_{min}) \
\frac{{\it\Gamma}_{M}^{0}}{({ E}-{ E}_{M}^{0})^{2} +
({{\it\Gamma}_{M}^{0}} / {2})^{2}}.    \label{omega-BW}
\end{equation}
There is
\begin{equation}
c_{1} = 1,\;\;\;c_{2} =  -\,2\,
\frac{ { E}_{0}\,-\,{E}_{min}}{({\it\Gamma}_{M}^{0})^{2}\;(\beta^{2} + \frac{1}{4}) }, \label{c1+c2}
\end{equation}
for $\omega (E) = \omega_{BW}  (E)$, (see \cite{R-U} for details). Here $\beta = \frac{E_{0} - E_{min}}{{\it\Gamma}_{M}^{0}}$.
In general, the sign of $c_{2}$ depends on the model considered (that is on the form of $\omega (E)$) and, contary to the case of $\omega (E) = \omega_{BW} (E)$, there is $c_{2} > 0$ for a wide class of $\omega (E) \neq \omega_{BW}(E)$.

The typical form of the survival probability ${\cal P}(t) = |A(t)|^{2}$ at transition times is shown  below in panel $A$ of  Fig (\ref{f1}) and Fig (\ref{f2}).
The behavior of ${\it\Gamma}_{M}(t)$ at
 canonical decay times $t < T_{1}$, at transition times $t \in(T_{1}, T_{2})$ and asymptotically late times $t > T_{2}$  is shown in panel $B$ of  Figs (\ref{f1}) and (\ref{f2}). These results
 are the direct, mathematical  consequence (by (\ref{h}) and (\ref{E=Re-h})) of properties of the amplitude $A(t)$ at these time regions.
 It is seen from these Figures that at times $t < T_{1}$, ${\it\Gamma}_{M}(t) \simeq {\it\Gamma}_{M}^{0}$ to a very high accuracy, then rapid and large fluctuations of ${\it\Gamma}_{M}(t)$ occur at the transitions time region $t \in (T_{1}, T_{2})$, and for very late times, $t > T_{2}$,  ${\it\Gamma}_{M}(t) \to 0 $ as $t \to \infty$ according to the result (\ref{G(t)-as}).

There is a widespread belief that the quantum theory accurately depicts reality.
This belief is based on the facts that
predictions of the quantum theory were confirmed experimentally to a very high accuracy.
So it should be expected with the  probability close to a certainty that
the experimental confirmation of the presence of  late time deviations from the exponential decay \cite{rothe}  means that
the late time properties of ${\it\Gamma}_{M}(t)$ and $E_{M}(t)$ described in Eqs (\ref{G(t)-as}) and (\ref{E(t)-as}) and
effects shown in  panel $(B)$ of Figs (\ref{f1}) and (\ref{f2}) should take place too, and should manifest itself under suitable conditions.
Results presented in Figs (\ref{f1}), (\ref{f2}) were obtained for
the  Breit--Wigner   energy distribution function (\ref{omega-BW})
assuming for simplicity that $\beta = 10$.

\section{Instantaneous energy $E_{M}(t)$ and the vacuum energy density at late times}

From the point of view of the purpose of the paper specified in the Introduction and the results presented in Sec. 3 the most important is the knowledge of the late time asymptotic properties of the decay rate. ${\it\Gamma}_{M}(t)$. Nevertheless, for the sake of completeness and for the convenience of readers, this Section will briefly discuss the late time asymptotic  properties of the energy $E(t)$ of an unstable system,  which can be applied to the analysis of the evolution of a universe having metastable vacuum.

As it was mentioned in Sec. 2 in \cite{Krauss} the idea was formulated that in the case of the metastable vacuum states some space regions or universes
can survive up to  times much later than times when the exponential decay law
holds. In the mentioned paper by Krauss and Dent the attention
was focused on the possible
behavior of the unstable false vacuum at very late times, where deviations from the exponential
decay law become to be dominat. In \cite{PRL-2011} it was concluded that such an effect
has to change the energy, $E_{M}^{0}$ of the system in the false (metastable) vacuum state at these times $t$ so that
at very long times $E_{M}^{0}$ is replaced by $E_{M}(t)$ and at these times the typical form of $E_{M}(t)$ looks as it results from the formula (\ref{E(t)-as}).

\begin{figure}[h!]
\begin{center}
{\includegraphics[width=75mm]{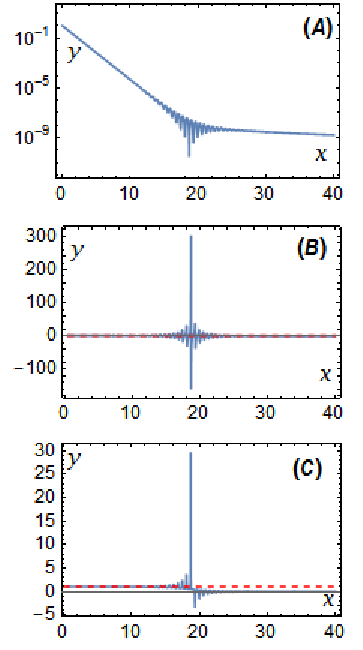}}
\caption{Typical form of the decay curve (panel $(A)$), the decay rate (panel $(B)$) and instantaneous energy (panel $(C)$) of an unstable state as a function of time. Axes: In all panels $x = t/\tau_{M}$ (The time $t$ is measured as a multiple of the lifetime $\tau_{M}$); Panel $(A)$ --- $y={\cal P}(t)$ (the logarithmic scale) --- the survival probability; Panel $(B)$ ---  $y= {\it\Gamma}_{M}(t) /  {\it\Gamma}^{0}_{M}$; Panel $(C)$ --- $\kappa (t)$ (The instantaneous energy in relation to the energy measured at canonical decay times). The horizontal dashed line $y=1$   represents in Panel $(B)$ the value of  $ {\it\Gamma}_{M}(t) /  {\it\Gamma}^{0}_{M} = 1$ whereas in Panel $(C)$ it represents $\kappa (t) = 1$}
  \label{f1}
  \end{center}
  \end{figure}

\begin{figure}[h!]
\begin{center}
{\includegraphics[width=75mm]{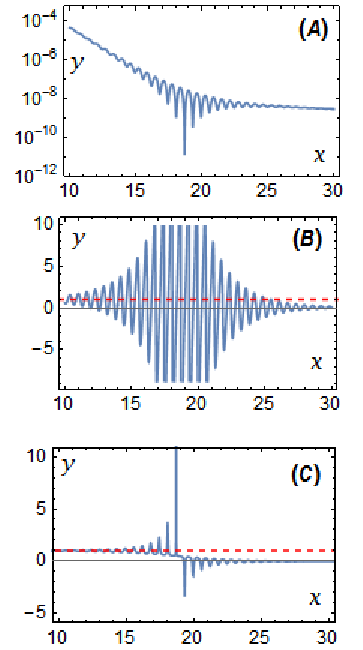}}
\caption{Enlarged  part of Fig (\ref{f1}) showing the behavior the survival probability ${\cal P}(t)$, decay rate ${\it\Gamma}_{M}(t)$ and $E(t)$ at the  of the transition times region.
Axes: $y= {\it\Gamma}_{M}(t) /  {\it\Gamma}^{0}_{M}$, $x= t/\tau_{M}$.  The horizontal dashed line $y=1$ represents in Panel $(B)$ the value of ${\it\Gamma}_{M}(t) \equiv  {\it\Gamma}^{0}_{M}$ whereas in Panel $(C)$ $\kappa (t) = 1$}
\label{f2}
  \end{center}
  \end{figure}

The typical behavior  of the energy $E_{M}(t)$ at
 canonical decay times $t < T_{1}$, at transition times $t \in (T_{1}, T_{2})$, (or $t \sim T$), and asymptotically late times $t >T_{2}$,
  is shown in  panels $(C)$ of  Figs (\ref{f1}), (\ref{f2}) (see also \cite{PLB-2014,Urbanowski:2016pks}) where the function
\begin{equation}
\kappa(t) = \frac{E_{M}(t) - E_{min}}{E^{0}_{M} - E_{min}} \label{kappa}
\end{equation}
is presented.
The red dashed line in these Figures denotes the value
\begin{equation}
\kappa (t) = 1,\;\;\; (t < T_{1}), \label{kappa=1}
\end{equation}
that is $E_{M}(t) = E^{0}_{M}$. Note that there is $E^{0}_{M} > E_{min}$. From these Figures it is seen that for $t < T_{1}$  we have $E_{M}(t) = E^{0}_{M}$ whereas for $t > T_{2} $  there is
\begin{equation}
E_{M}(t) - E_{min}  \simeq \pm\, c_{2}\, \frac{\hbar^{2}}{t^{2}},\;\; (t > T_{2}). \label{E_M-as}
\end{equation}

When one considers a meta--stable (unstable or decaying) vacuum state, $|M\rangle \equiv |0^{M}\rangle$,
the following important property of $\kappa (t)$ is useful:
\begin{eqnarray}
\kappa(t) &=& \frac{E_{M}(t) - E_{min}}{E^{0}_{M} - E_{min}} \nonumber\\
&\equiv &  \frac{\rho_{M}(t) - \rho_{bare}}{\rho^{0}_{M} - \rho_{bare}}, \label{kappa-vac}
\end{eqnarray}
where $\rho_{M} (t) = E_{M}(t)/V$ is the density of the vacuum energy in the decaying vacuum state considered,
$V$ is a volume,
$\rho^{0}_{M} = E^{0}_{M}/V$ is the density of the vacuum energy at times $t < T_{1}$,
$\rho_{bare} = E_{min}/V$ is the energy density in the true (bare) vacuum state, $|0^{\,bare}\rangle \equiv |0^{\,true}\rangle $, i.e., in the true ground state of the system.

From the last equations the following relation follows
\begin{equation}
\rho_{M}(t) - \rho_{bare} = (\rho^{0}_{M} - \rho_{bare})\,\kappa (t). \label{rho-M}
\end{equation}
Thus, because  for $t < T_{1}$ there is $\kappa (t) = 1$, one finds that
\begin{equation}
\rho_{M} (t) = \rho^{0}_{M},\;\;{\rm for}\;\; t <T_{1}, \label{rho-M-0}
\end{equation}
whereas for $t > T_{2}$ we have
\begin{equation}
\rho_{M} (t) - \rho_{bare} = (\rho^{0}_{M} - \rho_{bare})\,\kappa (t) \simeq \pm \,d_{2}\,\frac{\hbar^{2}}{t^{2}},\;\; (t \gg T). \label{rho}
\end{equation}
Analogous relations (with the same $\kappa (t))$ take place for $\Lambda (t) = \frac{8\pi G}{c^{2}}\,\rho_{M}(t)$.

The important property of $\kappa (t)$ is a presence of rapid fluctuations of the high amplitude for times $t \sim T$ , i. e. for $t \in (T_{1}, T_{2})$. This means that in the case of a decaying (unstable) vacuum analogous fluctuations of the energy density $\rho_{M} (t)$ and $\Lambda (t)$ should take place for $t \in (T_{1}, T_{2})$.
So if our Universe is the Universe with the unstable vacuum as the mass of Higgs boson suggests then in agreement with ideas expressed in \cite{Krauss}  we can conclude that the lifetime of the false  vacuum may be   shorter by at least a few or even much more orders than the age of our Universe. This means that
our Universe may place itself at the region of times described by the form of $\kappa (t)$ and ${\it\Gamma}_{M}(t)$ for $t > T_{2}$.

If one prefers to consider $\Lambda (t)$ instead of $\rho_{M} (t)$ then one obtains,
\begin{equation}
\Lambda (t) - \Lambda_{bare} = (\Lambda_{0} - \Lambda_{bare})\,\kappa (t), \label{lambda}
\end{equation}
or,
\begin{equation}
\Lambda (t) = \Lambda_{bare} +(\Lambda_{0} - \Lambda_{bare})\,\kappa (t), \label{lambda1}
\end{equation}
where $\Lambda_{0} = \frac{8\pi G}{c^{2}}\,\rho^{0}_{M}$ and $\Lambda_{bare} = \frac{8\pi G}{c^{2}}\,\rho_{bare}$.

One may expect that $\Lambda_{0}$ equals to  the cosmological constant calculated within quantum field theory \cite{sz1}. From (\ref{lambda1})
it is seen that for $t < T_{1}$,
\begin{equation}
 \Lambda_{M} (t) \simeq \Lambda_{0}, \;\;\;{\rm for}\;\;\;(t < T_{1}),
 \end{equation}
 because $\kappa (t < T_{1}) \simeq 1$. Now if to assume that $\Lambda_{0}$ corresponds to the value of the cosmological constant $\Lambda$ calculated within the quantum field theory, than one should expect that \cite{sz1}
 \begin{equation}
 \frac{\Lambda_{0}}{\Lambda_{bare}} \geq 10^{120}, \label{lambda2}
 \end{equation}
 (see \cite{sz1}) which allows one to write down Eq. (\ref{lambda1}) as follows
 \begin{equation}
 \Lambda_{M}(t) \simeq \Lambda_{bare} + \Lambda_{0}\,\kappa(t). \label{lambda3}
 \end{equation}
 Note that for $t >  T_{2}$ there should be (see (\ref{rho}))
 \begin{equation}
\Lambda_{0}\, \kappa (t) \simeq \frac{8\pi G}{c^{2}}\,d_{2} \frac{\hbar^{2}}{t^{2}},\;\;{\rm for}\;\; ( t > T_{2}), \label{lambda4}
 \end{equation}
that is
\begin{equation}
 \Lambda_{M}(t) \simeq \Lambda_{bare} \pm \frac{{\mathfrak{ b}}_{2}}{t^{2}},  \;\;{\rm for}\;\; ( t >  T_{2}),  \label{lambda5}
 \end{equation}
where ${\mathfrak{ b}}_{2} = \frac{8\pi G}{c^{2}}\,\hbar^{2}\,d_{2}$ and the sign of ${\mathfrak{b}}_{2}$ is determined by the sign of $d_{2}$.

Note that
a parametrization following from the quantum theoretical treatment of meta\-stable vacuum states can explain why the cosmologies with the time--depen\-dent cosmological constant $\Lambda (t)$ are worth  considering and may help to explain the cosmological constant problem \cite{Weinberg,Caroll}. The time dependence of  $\Lambda$ of the type $\Lambda (t) = \Lambda_{bare} + \frac{\alpha^{2}}{t^{2}}$ was assumed eg. in \cite{Lopez} but there was no any explanation what
physics suggests such a choice of the form of $\Lambda$. Earlier analogous form of $\Lambda$ was obtained in \cite{Canuto}, where  the  invariance under scale transformations of the generalized Einstein equations was studied. Such a time dependence of $\Lambda$ was
postulated also in \cite{Lau} as the result of the analysis of the large numbers hypothesis. The cosmological model with  time dependent $\Lambda$ of the above postulated form was studied also in \cite{Berman} and in much more recent papers.

The nice feature and maybe even the advantage of the
formalism presented in Section 4 and
in this Section
is that in the case of the universe with metastable (false) vacuum if one realizes that the decay of this unstable vacuum state is the quantum decay process then it
emerges automatically that there have to exist the true ground state of the system that is the true (or bare)  vacuum with the minimal energy, $E_{min}> - \infty$, of the system corresponding to this vacuum, and equivalently, $\rho_{bare} = E_{min}/V$, or $\Lambda_{bare}$.
What is more, in this case the  $\Lambda \equiv \Lambda(t)$ having the form described by equations  (\ref{lambda3}) --- (\ref{lambda5}) emerges quite naturally.
In such a case   the function $\kappa(t)$ given by the relation (\ref{kappa}) describes time dependence for all times $t$ of the energy density $\rho_{M}(t)$ (or the cosmological "constant" $\Lambda_{M}(t)$) and its general form is presented in panels $(C)$ in Figs (\ref{f1}) and (\ref{f2}).
Note that results presented in Sections 4 and 5 are rigorous.

As mentioned in the introduction to this Section, its aim is to inform readers about  the late--time properties of the energy density  $\rho_{M}(t)$ in the  false vacuum state and how they can affect on the behavior of $\rho_{M}(t)$ and $\Lambda_{M}(t)$ at late times (see Eqs (\ref{rho}), and  (\ref{lambda3}), (\ref{lambda4})).
We do not present here a more detailed analysis of a possible cosmological consequences of these properties because
detailed discussion and analysis of the  consequences of the late time properties of the density of the vacuum energy $\rho_{M} (t)$ and $\Lambda (t)$  briefly described in this Section readers can find in \cite{Urbanowski:2016pks,sz1,ms+ku1,ms+ku2,epjc-2017b,jcap-2020}.

\section{Discussion and conclusions}

The problem how the process of an expansion of a universe and its decay process affect together on this universe is analyzed in Sec. 3. The possible result of these combined processes is characterized by the condition (\ref{main}).
 The obvious next step in the considerations in Sc. 3 and 4 is to apply the results obtained in them to the analysis of the possible future fate of the universe with an unstable vacuum.
 In the case of very late times
assuming that the lifetime of the metastable false vacuum is  shorter by at least a few or even much more orders than the age of our Universe
it can be done by
inserting into (\ref{main}), eg. the present value of the Hubble expansion rate $H(t) = H(t_{0}) = H_{0}$ and  the late time asymptotic form of the decay rate ${\it\Gamma}_{M}(t)$ given by relations (\ref{G(t)-as}) and (\ref{c-i}),
 \begin{equation}
{\it\Gamma}_{M} (t) \simeq 2(\lambda + 1)\,\frac{\hbar}{t}  \;\;\;({\rm for}
\;\;t > T_{2}), \label{G(t)-1}
 \end{equation}
where the  the coefficient $c_{3}$ in (\ref{G(t)-as}) is neglected, and  assuming that $t=t_{0}$, (where $t_{0}$ is the age of the our Universe). The only problem is to choose the appropriate value of $\lambda$ in (\ref{c-i}). If to choose $\lambda$ appearing in the case of the decays into two particles, that is, $\lambda = 1/2$ (see, eg. \cite{goldberger}) , then inserting  the present values of $H_{0}$ and $t_{0}$ \cite{pdg-2020} into (\ref{main}) one obtains that
\begin{equation}
d_{H,{\it\Gamma}_{M}} = 3H_{0} - \frac{{\it\Gamma}_{M}(t_{0})}{\hbar} \simeq 3\left[H_{0} - \frac{1}{t_{0}}\right], \label{d-HG-0}
\end{equation}
where $H_{0} = H(t_{0})$ is the present--day value of the Hubble parameter \cite{pdg-2020}, which gives
\begin{equation}
d_{H,{\it\Gamma}_{M}} \simeq - 0.001065\; [{\rm Gyr}]^{-1} \simeq -  3.3764\times 10^{-19}\;[{\rm s}]^{-1} < 0. \label{d-HG-1}
\end{equation}
This result suggests that in  the case  $\lambda = 1/2$ the Universe  may decay at  late times,
 but such a conclusion can not be considered  to be decisive and final. First, taking into account the neglected term $c_{3}\, \frac{\hbar^{3}}{t^{3}}$ in (\ref{G(t)-1}) can result in changing the sing of $d_{H,{\it\Gamma}_{M}}$. The second,
 there is no certainty that the choice $\lambda = 1/2$ is the correct choice. In fact, it is not known what value of $\lambda$ is correct for decays of the  unstable vacuum states and this problem requires further studies. So,
we need some bounds for the values of $\lambda$ that lead to the nonnegative $d_{H,{\it\Gamma}_{M}}$.
 The solution of the equation
\begin{equation}
d_{H,\it\Gamma} \stackrel{\rm def}{=}3\,H(t)\,-\,\frac{2(\lambda +1)}{t} = 0, \label{d-HG-2}
\end{equation}
which follows  from (\ref{main}) and (\ref{G(t)-1}) is
\begin{equation}
\lambda \simeq 0.426546. \label{lambda-s}
\end{equation}
This solution is obtained for the same values of $H(t)=H_{0}$ and $t=t_{0}$ which were used to find the result (\ref{d-HG-1}). The result (\ref{lambda-s}) means that there should be
\begin{equation}
d_{H,\it\Gamma}\; \geq 0 \;\;\;{\rm for}\;\;\; 0 \leq \lambda \leq 0.426546 \label{d_HG>0}
\end{equation}
and
\begin{equation}
d_{H,\it\Gamma}\; < 0 \;\;\;{\rm for}\;\;\; \lambda > 0.426546, \label{d-HG<0}
\end{equation}
within the considered late time approximation (\ref{G(t)-1})  for ${\it\Gamma }(t)$.
Thus, if the energy distribution $\omega (E)$ for the universe in the metastable vacuum state is given by the relation (\ref{omega-eta}) with such $\lambda$
that $0 \leq \lambda \leq 0.426546$, then  such a universe  should rather not decay. This conclusion show how important is to find $\omega (E)$ and thus $\lambda$ for the metastable vacuum state of the universe. To complete this discussion let us note that the Breit--Wigner energy distribution function (\ref{omega-BW}) corresponds to the  case $\lambda = 0$.
This means that in the considered case of the late times, when the late time approximations for $\Lambda (t)$ and ${\it\Gamma}_{M}(t)$ hold,
the use of the Breit--Wigner form of $\omega (E)$ to characterize the false vacuum state can give our Universe stability.
Unfortunately, it is not certain currently whether such $\omega(E)$ correctly characterizes the energy distribution density in the false vacuum state.
Among other things, for this reason, it is necessary to study  properties of the metastable false vacuum state  and the corresponding $\omega(E)$.
As it is seen from the results presented in Sec.4, the coefficients $c_{1}, c_{2}, \ldots,$ in late time asymptotic expansions of ${\it\Gamma}(t)$ and $E_{M}(t)$ depend on the form of $\omega (E)$ (see Eq. (\ref{c1+c2})). Therefore simply
the knowledge of the correct $\omega (E)$ is necessary when one wants to find the proper  form, values and sign of the coefficient $c_{2}$ appearing in relations (\ref{E(t)-as})   and (\ref{E_M-as}) and then $d_{2}$ in (\ref{rho}), and also $c_{3}$ in (\ref{G(t)-as}),
 but above all, knowing the correct $\omega (E)$ we will be able to answer the question of whether the hypothesis mentioned in Sec. 1 and formulated  by Krasus and Dent in \cite{Krauss} is realized in our Universe

One may ask what do the results presented in this paper really mean? Suppose that our Universe was created in the metastable false vacuum state and  the lifetime of this vacuum is much shorter than the time $T_{2}$ defined in Sec. 2 by Eq. (\ref{T2}) and that this $T_{2}$ is much shorter  than the age of the Universe.
Then in our epoch its survival probability, ${\cal P}(t_{0})$, is negligibly small: One can even say that it is   zero to very high accuracy. The methods used in this paper and the quantum theory of unstable states do not give an answer for the question when this Universe should decay but they can explain why this Universe still exists and whether  will it exist longer?  In the light  ideas presented in,  e.g. \cite{Esp1} and in other papers mentioned in Sec. 1, such an information seems to be very important.

Note that formalism and results described  in Sections 2, 4 and 5 are rigorous.
The approach described in Sections 2 and 5 was  applied
in \cite{Urbanowski:2016pks,sz1,ms+ku1,ms+ku2,epjc-2017b,jcap-2020}, where cosmological models with $\Lambda(t) = \Lambda_{bare} \pm \frac{\alpha^{2}}{t^{2}}$ were studied.
From the results presented therein and in this paper, in the light of  the LHC result concerning the mass of the Higgs boson \cite{pdg-2020} and its cosmological consequences, the conclusion follows that further studies of this approach are necessary.

\hfill\\
{\bf Funding:} This research received no external funding.

\hfill\\
{\bf The author contribution statement:}   The author confirms and  assures that he  is the sole author of the article. The author declares that there are no conflicts of interest
regarding the publication of this article  and that all results presented in this article are the author's own results.

\


\begin{thebibliography}{20}
\bibitem{Coleman}
S. Coleman, \emph{{The fate of the false vacuum. 1. Semiclassical theory}}, Phys. Rev. D 15, 2929 (1977).
\bibitem{Callan}
C.G. Callan and S. Coleman,  \emph{{The fate of the false vacuum. 2.
  First quantum corrections}}, Phys. Rev. D 16, 1762 (1977).
\bibitem{Col2}
S. Co1eman and T.  De Luccia, {\em Gravitational effects on and of vacuum decay}, Physical Review D, {\bf 21}, 3305--- 3315, (1980).
\bibitem{Krauss}
L. M. Krauss, J. Dent, \emph{{The late time behavior of false vacuum decay:
  Possible implications for cosmology and metastable inflating states}}, Phys. Rev. Lett., \textbf{100}, 171301 (2008).
\bibitem{Winitzki}
S. Winitzki, \emph{Age-dependent decay in the landscape}, Phys. Rev. {\bf D 77}, 063508 (2008).

\bibitem{Ade}
Ade, P. A. R., Ahmed, Z., Aikin, R. W., Alexander, K. D., Barkats, D., Benton, S.
J., et al. (2012).{\em  Observation of a new particle in the search for the Standard
Model Higgs boson with the ATLAS detector at the LHC}, Phys. Lett. B 716,
1–29. doi: 10.1016/j.physletb.2012.08.020


\bibitem{Chat}
Chatrchyan, S., et al. (2012). {\em Observation of a new boson at a mass of 125
GeV with the CMS experiment at the LHC}, Phys. Lett. B 716, 30–61.
doi: 10.1016/j.physletb.2012.08.021


\bibitem{pdg-2020}
P.A. Zyla et al. ({\em Particle Data Group}), Prog. Theor. Exp. Phys., {\bf  2020}, 083C01 (2020);
doi: 10.1093/ptep/ptaa104


\bibitem{Deg}
Degrassi, G., Di Vita, S., Elias-Miro, J., Espinosa, J. R., Giudice, G. F., Isidori, G.,
et al. (2012). {\em Higgs mass and vacuum stability in the StandardModel at NNLO},
J. High Energy Phys. 8:98 (2012); doi: 10.1007/JHEP08(2012)098


\bibitem{But}
Buttazzo, D., Degrassi, G., Giardino, P. P., Giudice, G. F., Sala, F., Salvio, A., et al.
(2013). {\em Investigating the near-criticality of the Higgs boson}, J.High Energy Phys.
12:89. doi: 10.1007/JHEP12(2013)089

\bibitem{Isidori}
G. Isidori, R. Ridolf, A. Strumia, {\em  On the metastability
of the standard model vacuum}, Nucl. Phys. {\bf B 609}, 387, (2001).


\bibitem{bamba}
K. Bamba, S. Capozziello, S. Nojiri and S. D. Odintsov, \emph{Dark energy cosmology: the equivalent description via different
theoretical models and cosmography tests,}
Astrophys. Space Sci.  {\bf 342}, 155, (2012);
[arXiv:1205.3421 [gr-qc]].


\bibitem{graef}
Elcio Abdalla, L. L. Graef, Bin Wang, \emph{A model for dark energy decay}, Physics Letters {\bf B726}, 786 --- 790, (2013).


\bibitem{Spencer}
 A. Kobakhidze, A. Spencer--Smith,  {\em Electroweak vacuum (in)stability in an inflationary universe}, Phys. Lett. {\bf B 722}, 130, (2013).



\bibitem{Kob}
A. Kobakhidze and A. Spencer--Smith, {\em The Higgs vacuum is unstable}, arXiv:1404.4709v2 [hep-ph], (2014);
doi: 10.48550/arXiv.1404.4709

\bibitem{Esp}
José R. Espinosa, {\em Implications of the top (and Higgs) mass for vacuum
stability}, (8th International Workshop on Top Quark Physics, TOP2015
14---18 September, 2015, Ischia, Italy),  Proceedings of Science, (TOP2015) 043.

 \bibitem{Elias}
 J. Elias--Miro, J. R.  Espinosa, G. F.  Giudice, G.  Isidori, A.  Riotto, A., and A. Strumia,
A. (2012). Higgs mass implications on the stability of the electroweak vacuum.
Phys. Lett. {\bf B 709}, 222--228. (2012); doi: 10.1016/j.physletb.2012.02.013


\bibitem{Wei}
Wei Chao, Matthew Gonderinger, and Michael J. Ramsey--Musolf, {\em Higgs vacuum stability, neutrino mass, an  dark matter}.
Phys. Rev. {\bf D 86}, 113017, (2012).

\bibitem{Ema}
Yohei Ema, Kyohei Mukaida and Kazunori Nakayama, {\em Fate of electroweak vacuum during
preheating}, Journal of Cosmology and Astroparticle Physics, {\bf 10}, 043, (2016); doi:10.1088/1475-7516/2016/10/043

\bibitem{Esp1}
J. R. Espinosa, {\em et al}, \emph{The cosmological Higgstory of the vacuum instability}, JHEP 09, 174 (2015); doi:10.1007/JHEP09(2015)174; ArXiv: 1505.04825.



\bibitem{Mar}
T. Markkanen, A. Rajantie and S. Stopyra, {\em Cosmological Aspects of Higgs
Vacuum Metastability},
Frontiers in Astronomy and Space Sciences, {\bf 5}, Article No: 40, (2018);
doi: 10.3389/fspas.2018.00040

\bibitem{Stojk2}
D.~C.~Dai, R.~Gregory and D.~Stojkovic,
{\em Connecting the Higgs Potential and Primordial Black Holes},
Phys. Rev. D \textbf{101}, 125012 ,(2020);
doi:10.1103/PhysRevD.101.125012;
[arXiv:1909.00773 [hep-ph]]


\bibitem{JK}
John Kearney, Hojin Yoo and Kathryn M. Zurek, \emph{Is a Higgs vacuum instability fatal for high-scale inflation?}, Physical Review D {\bf 91}, 123537 (2015);
doi: 10.1103/PhysRevD.91.123537.

\bibitem{De-Ch}
De--Chang Dai, Ruth Gregory and Dejan Stojkovic, \emph{Connecting the Higgs potential and primordial black holes}, Physical Review, {\bf D 101}, 125012, (2020);
doi: 10.1103/PhysRevD.101.125012.

\bibitem{Bur1}
Philipp Burda, Ruth Gregory and Ian G. Moss,  \emph{Vacuum metastability with black holes}, JHEP 08,  114 (2015); doi:10.1007/JHEP08(2015)114; ArXiv: 1503.07331.

\bibitem{Bur2}
Philipp Burda, Ruth Gregory and Ian G. Moss, \emph{The fate of the Higgs vacuum}, JHEP 06, 025  (2016); doi:10.1007/JHEP06(2016)025; ArXiv: 1601.02152.


\bibitem{Branchina1}
V.~Branchina and E.~Messina, \emph{{Stability, Higgs boson mass and new
  physics}},
   {Phys. Rev. Lett.} {\bfseries 111},  241801, (2013).

\bibitem{Branchina2}
V.~Branchina, E.~Messina and M.~Sher, \emph{{Lifetime of the electroweak vacuum
  and sensitivity to Planck scale physics}},
  {{Phys. Rev.}
  {\bfseries D91}, 013003,  (2015)}.


\bibitem{bohm}
A. Bohm, Quantum Mechanics: Foundations and Applications, $2^{\rm nd}$--ed., Springer 1986.



\bibitem{Fock}
S. Krylov, V. A. Fock,  \emph{{On two main interpretations of energy-time
  uncertainty}}, Zh. Eksp. Teor. Fiz. {\bf 17}, (1947), 93.
\bibitem{Fonda}
L. Fonda, G. C. Ghirardii and A. Rimini, \emph{{Decay Theory of Unstable Quantum
  Systems}}, Rep. on Prog. in Phys. {\bf 41},  587 (1978).

\bibitem{Khalfin}
L. A. Khalfin, \emph{{Contribution to the decay theory of a quasi-stationary
  state}}, Zh. Eksp. Teor. Fiz. {\bf 33}, (1957), 1371 [Sov.
Phys. --- JETP {\bf 6}, (1958), 1053].


\bibitem{Paley}
R. E. A. C. Paley, N. Wiener, {\em Fourier
transforms in the complex
domain}, American Mathematical Society, New York, 1934.





\bibitem{WW}
V. F. Weisskopf, E. T. Wigner, emph{Berechnung der nat\"{u}rlichen Linienbreite auf Grund der Diracschen Lichttheorie},
Z. Phys. \textbf{63}, 54, (1930);
\emph{\"{U}ber die nat\"{u}rliche Linienbreite
in der Strahlung des harmonischen Oszillators}, Z. Phys.
 \textbf{65}, 18, (1930).


\bibitem{rothe}
C. Rothe,
S. I. Hintschich and A. P. Monkman, \emph{Violation of the Exponential-Decay Law at Long Times}, Phys. Rev. Lett. {\bf 96}, 163601 (2006).

\bibitem{pra}
K. Urbanowski, \emph{{Early-time properties of quantum evolution}}, Phys. Rev. \textbf{ A 50},  2847, (1994).


\bibitem{Cheng}
Ta--Pei Cheng, \emph{Relativity, Gravitation,
and Cosmology:
A basic introduction}, Oxford University Press, 2005.

\bibitem{Sahni}
V. Sahni and A. Starobinsky, \emph{The Case for a Positive Cosmological
$\Lambda$--term},  International Journal of Modern
Physics D, {\bf  09}, 373 --- 443, (2000).


\bibitem{urbanowski-2-2009}
K. Urbanowski, \emph{Long time properties of the evolution of an unstable state}, Cent. Eur. J. Phys. {\bf 7},  696, (2009);
DOI:10.2478/s11534--009--0053--5.

\bibitem{urbanowski-1-2009}
K. Urbanowski, \emph{{General properties of the evolution of unstable states at
  long times}}, Eur. Phys. J. {\bf D 54}, 25,  (2009).


 \bibitem{giraldi}
 F. Giraldi, \emph{{Logarithmic decays of unstable states}}, Eur. Phys. J. D, {\bf 69}: 5 (2015).

\bibitem{PLB-2014}
K. Urbanowski and K. Raczy\'{n}ska, \emph{Possible emission of cosmic X-- and $\gamma$--rays by unstable particles
at late times},
Phys. Lett. {\bf B 731}, 236 (2014).


\bibitem{R-U}
K.~Raczy\'{n}ska and K.~Urbanowski, \emph{Survival amplitude, instantaneous energy
  and decay rate of an unstable system: Analytical results},
  Acta Phys. Polon.
  {\bf B49}, 1683, (2018).










\bibitem{PRL-2011}
K. Urbanowski,  \emph{{Comment on "Late time behavior of false vacuum decay:
  Possible implications for cosmology and metastable inflating states"}},
Phys. Rev. Lett., {\bf 107}, 209001 (2011).



\bibitem{Urbanowski:2016pks}
K.~Urbanowski, \emph{{Properties of the false vacuum as the quantum unstable
  state}},
  {{Theor. Math., 458,
  Phys.} {\bfseries 190} (2017).}





\bibitem{sz1}
M. Szyd{\l}owski,
\emph{Cosmological model with decaying vacuum energy from quantum mechanics},
Phys. Rev. {\bf D 91}, 123538 (2015).



\bibitem{Weinberg}
S. Weinberg, \emph{{The cosmological constant problem}},
Rev. Mod. Phys. {\bf  61}, 1 (1989).




\bibitem{Caroll}
S. M. Carroll, {\em The Cosmological Constant}, { Living Rev. Relativity}, 3, (2001), 1;
http://www.livingreviews.org/lrr-2001-1.





\bibitem{Lopez}
J. L. Lopez and D. V. Nanopoulos, \emph{ A new cosmological constant model},
Mod. Phys. Lett. {\bf A 11}, 1
(1996).


\bibitem{Canuto}
V. Canuto and S. H. Hsieh,
\emph{Scale--Covariant Theory of Gravitation and Astrophysical Applications },
Phys. Rev. Lett. {\bf 39}, 429, (1977).

\bibitem{Lau}
Y. K. Lau and S. J. Prokhovnik,
\emph{The larg numbr hyphotesis and a relativistic theory of gravitation},
Aust. J. Phys., {\bf 39}, 339, (1986).

\bibitem{Berman}
M. S. Berman,
\emph{Cosmological models with a variable cosmological term },
Phys. Rev. {\bf D 43}, 1075, (1991).









\bibitem{ms+ku1}
K. Urbanowski and M. Szyd{\l}owski,
\emph{Cosmology with a decaying vacuum},
AIP Conf. Proc. {\bf 1514}, 143 (2013).


\bibitem{ms+ku2}
M Szyd{\l}owski, A Stachowski and K Urbanowski,
\emph{Cosmology with a decaying vacuum energy
parametrization derived from quantum mechani},
Journal of Physics: Conference Series {\bf 626},  012033, (2015).




\bibitem{epjc-2017b}
A. Stachowski1, M. Szydłowski1, K. Urbanowski,
\emph{Cosmological implications of the transition from the false vacuum
to the true vacuum state},  { Eur. Phys. J.} {\bf  C  77}, 357, (2017).


\bibitem{jcap-2020}
M. Szyd{\l}owski, A. Stachowski
and K.  Urbanowski, \emph{The evolution of the FRW universe with decaying metastable dark energy --- a dynamical system analysis},
{ Journal of Cosmology and Astroparticle Physics},
{\bf 04} 029 (2020).




\bibitem{goldberger}
M. L. Goldberger, K. M. Watson, {\em Collision theory}, Wiley, New York, 1964.


\bibitem{ku-2022}
K. Urbanowski, \emph{Cosmological "constant" in a universe born in the metastable false
vacuum state}, Eur. Phys. J. C,   82:242 (2022);
https://doi.org/10.1140/epjc/s10052-022-10195-2

\end{thebibliography}
\end{document}